# Darkflight estimates of meteorite fall positions: issues and a case study using the Murrili meteorite fall

M.C. Towner[1,2], T. Jansen-Sturgeon[2], M. Cupak[2], E.K. Sansom[2], H.A.R. Devillepoix[2], P.A. Bland[2], R.M Howie[2], J.P. Paxman[2], G.K. Benedix[2], B.A.D. Hartig[2].

Abstract

Fireball networks are used to recover meteorites, with the context of orbits. Observations from these networks cover the bright flight, where the meteoroid is luminescent, but to recover a fallen meteorite, these observations must often be predicted forward in time to the ground to estimate an impact position. This darkflight modelling is deceptively simple, but there is hidden complexity covering the precise interactions between the meteorite and the (usually active) atmosphere. We describe the method and approach used by the Desert Fireball Network, detailing the issues we have addressed, and the impact that factors such as shape, mass and density have on the predicted fall position. We illustrate this with a case study of Murrili meteorite fall that occurred into Lake Eyre-Kati Thanda in 2015. The fall was very well observed from multiple viewpoints, and the trajectory was steep, with a low altitude endpoint, such that the darkflight was relatively short. Murrili is 1.68 kg with a typical ordinary chondrite density, but with a somewhat flattened shape compared to a sphere, such that there are discrepancies between sphere-based predictions and the actual recovery location. It is notable that even in this relatively idealised darkflight scenario, modelling using spherical shaped projectiles resulted in a significant distance between predicted fall position and recovered meteorite.

[1] Corresponding author

[2] Space Science and Technology Centre, Curtin University, GPO Box U1987, Perth, WA 6845, Australia

# 1 INTRODUCTION

Camera networks are used to observe fireballs, for the study of meteoroid orbits, and the recovery of meteorites with known orbits. By recording the arrival of a fireball one can calculate the arrival trajectory and hence the orbit and origin within the solar system. If the fireball is large enough that a meteorite falls, recovery of this is often a high scientific priority, as it represents a fresh solar system sample with a known origin. Several networks have been deployed historically, beginning with the Harvard photographic meteor program (Jacchia and Whipple 1956), with the first observed and recovered meteorite being Příbram from the Czech Fireball Network in 1959 (Ceplecha 1961).

Initially these systems were film-based but more recently digital systems have become predominant (for example, (Spurný et al. 2006; Colas et al. 2014; Howie et al. 2017)). The practical techniques they operate upon is to observe fireballs from multiple viewpoints, and then triangulate these bright observations to derive a trajectory in the atmosphere (Ceplecha 1987; Borovicka 1990), and hence backtrack to calculate a heliocentric orbit. In the case of a putative meteorite, for recovery, one uses this bright flight trajectory to calculate forward in time to give a predicted fall position on the ground; the so-called darkflight analysis, allowing ground searches to then be carried out. The details of this darkflight calculation are not often discussed in the literature, except in passing that usually imply that integration calculations were carried out. The calculation is done using the classical drag equation (Equation 1), but this simplicity hides more complex factors. For the starting conditions, what is the shape, mass, density of the meteorite? How do these properties reflect in factors like the drag coefficient? What is the behaviour of the atmosphere at that time and position? How are uncertainties propagated? Typically, bright flight end points are 20-30 km altitude, so there is significant height and time that must be integrated through to get to the ground, and small errors can accumulate resulting in significant errors in the predicted fall position compared to actual landing site. This darkflight calculation is also difficult to verify. By the very nature of darkflight, there are no observations to cross check during descent, so the only criteria for successful modelling is location and characteristics of a recovered meteorite, and a failure to recover may be caused by factors unrelated to the darkflight.

## 1.1 Historical review

In literature concerning meteorite recoveries, the precise method used for darkflight prediction is discussed, but rarely in full detail. One early example given in (Ceplecha 1961), describing the Příbram meteorite fall, where the winds were a special case of directly against the azimuth of meteor trajectory, simplifying calculations. This predates the widespread use of computers, so calculations were integrated numerically by hand, which the authors describe as 'laborious'.

Integration steps were every 100 m altitude, and drag coefficient was a function of Mach number, but fixed in subsonic regime to a spherical value of 0.52.

(Ceplecha 1987) (see also (Ceplecha et al. 1998) p390) described their approach in some detail, following the discussion of triangulation methods, as an integration under atmospheric interactions and gravity. They used a Runge-Kutta integrator with a fixed integration step of 10 m (so in space rather than in time) and assumed that after the apparent end of the luminous phase, ablation has ceased and there is no fragmentation, such that meteorite shape does not change. Nearest observational data was used for atmospheric density and wind values, supplemented by a standard atmosphere model (CIRA 1972) if observations were incomplete. Drag coefficient was a function of Mach number (unspecified), with a scalar modifier for a non-spherical nature fixed by using the deceleration at the end point of bright flight. This scalar parameter is in contrast to their work focused on bright flight (Pecina and Ceplecha 1983; Ceplecha et al. 2000) where photometric mass is introduced and used to separate mass from a shape density scalar.

Further publications on notable recovered meteorites briefly mention details of their approaches to darkflight: For the Lost City meteorite, (McCrosky et al. 1971) carry out a numerical integration assuming a spherical rock, using a standard atmosphere, with a fixed drag coefficient of a sphere of 0.92 (using a different nomenclature to 0.52 number for (Ceplecha 1961)). For the Innisfree meteorite recovery, (Halliday et al. 1978) note that they use atmospheric data from the nearest balloon flight observations at Edmonton. (Gritsevich et al. 2014) describe the Annama meteorite observations and recovery, with a Monte Carlo simulation of the darkflight using wind data provided by the Finnish Meteorological Institute. They describe their method in some detail, and begin by fitting the later part of the observed bright flight, calculating forwards and requiring any fragments to have an appropriate shape/density/mass to fit subsequent bright flight observations. For the Desert Fireball Network (DFN) recovery of Bunburra Rockhole (Spurný et al. 2012), the paper summarises the DFN approach as applied at the time briefly, here we describe the current DFN approach in more detail.

Some studies have also focused on the drag coefficient parameter; (Zhdan et al. 2007) describe the drag coefficient of various meteorite shapes, as does (Carter 2011). Additionally, there are many studies on the general aerodynamic drag coefficients for hyper/super/subsonic objects, that we refer to in later sections.

## 2 METHOD

Here we describe the darkflight modelling technique used by the Desert Fireball Network, discussing the details of some factors that affect the predictions. We illustrate this with case study of the Murrili fireball.

### 2.1 Overview

The DFN data pipeline is almost completely automated, so it would be in principle possible to generate WRF and darkflight predictions for all triangulations seen. However, such an approach would overwhelm, without some method of filtering. Hence, the choices of which fireballs to investigate in detail is based primarily on the a-b criteria from (Gritsevich 2007; Sansom et al. 2019). Further heuristics that are often considered are considerations of the final observed bright velocity and end height.

For darkflight modelling, one starts with the classical aerodynamic drag equation:

$$\underline{F_d} = -\frac{C_d \rho S v_{mag}^2 \underline{v}}{2m},$$

Equation 1

where $\underline{F_d}$ is the drag force on the body, $C_d$ is the drag coefficient, m is the body mass, S is the body cross-sectional area, $\underline{v}$ is the unit velocity vector relative to the atmosphere (which includes any contribution from wind movement of the atmosphere) with a magnitude of $v_{mag}$, and $\rho$ is the atmospheric mass density. Gravitational forces must also be calculated and included for a trajectory calculation.

This equation is used computationally; one integrates forward in time, initiating parameters from the last observed bright flight position, consisting of position, velocity vector, meteorite mass and shape. Figure 1 outlines the calculation steps involved. For every position, the appropriate environmental conditions (air pressure, temperature, density (or humidity), wind speed/direction, gravity vector) are either calculated or looked up in data tables. It is then possible to calculate the Knudsen number, the Reynolds number, and hence the drag coefficient throughout the descent of the body (see Table 1). Hence we calculate forces on the body, accelerations, and update position and velocity vectors using Newtonian mechanics, and account for any ablation effects that will change the mass and hence cross-sectional area, using a theoretical estimation (since there are no observations). For starting conditions, one can also include derived parameters from theoretical modelling of the bright flight behaviour: mass, meteorite shape, and plausible density. One can also impose further conditions that the transition from bright to dark flight must be smooth; most importantly this means that the rate of change of acceleration should be smooth from bright to dark flight (c.f. (Ceplecha 1987)). The DFN approach for the core procedure is monotonic; we assume shape does not change during descent, and that there is no fragmentation (c.f. (Vinnikov et al. 2016)). In reality, it is quite possible for fragmentation to occur during the darkflight, as seen in meteorites recovered with broken or missing fusion crust (Folinsbee and Bayrock 1961; Spurný et al. 2012). Only a small amount of ablation is predicted between the time after the meteorite ceases to be observed and the time where the velocity has dropped sufficiently for ablation to actually cease, and we implement this in code using the approximation of (Passey and Melosh 1980) equation 2. Typical values for this later ablation are predicted to be <1% of the final mass, which is a relatively small uncertainty compared to other factors.

In common with previously described methods for darkflight integration, the core of the DFN approach is a time series integration. We have tested simple first order time step integration but get slightly better fidelity (in terms of matching fall positions of recovered meteorites) using a 4th order Runge-Kutta integrator.

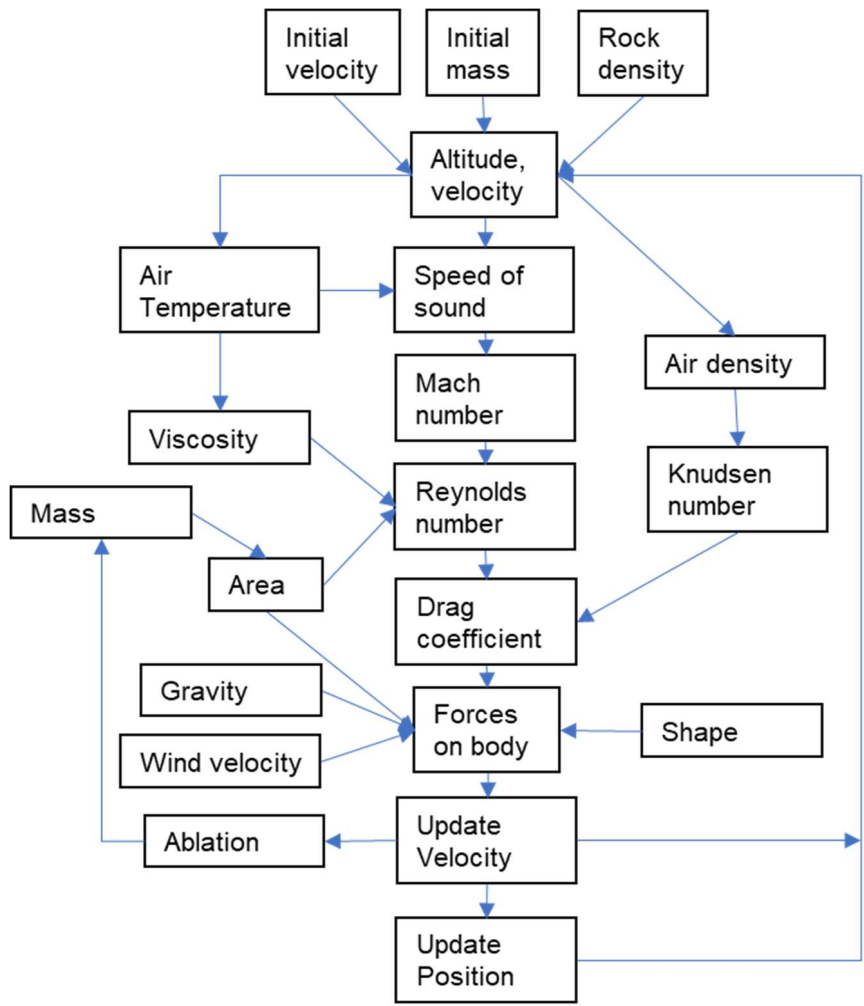

*Figure 1 – flow schematic of darkflight modelling.*

## 2.2 Complicating Factors

Darkflight integration presents a unique testing problem, as there are no observations during descent that can offer insight as to the accuracy of the approach. The only 'test' available is the recovery—or not—of the meteorite, and its final position with respect to the prediction. To compound this, non-recovery of a meteorite may not indicate errors in darkflight modelling, as it may be a ground searching issue. The only other assistance one may get is serendipitous, non-visual observations; in the case of large meteorite, weather Doppler radar may detect the falling body or bodies (Fries and Fries 2010; Jenniskens et al. 2012), or if a seismic sensor is very close by, the impact may be detected. A specialist instrument such as an active RADAR or LIDAR would assist, but typical ranges are relatively small (few 10s of km) and such instruments are expensive and have not been deployed to date, to the authors knowledge.

From theoretical modelling such as (Sansom et al. 2017) or earlier work such as (ReVelle 2005) and references therein, the shape, density and mass are only partially constrained and are interlinked. However, one can apply some plausible constraints; it is reasonable to expect the density to be close to one of three typical values – approximately 2700 kg/m$^3$ for an achondrite, 3500 kg/m$^3$ for a chondrite, or 7500 kg/m$^3$ for iron (Flynn 2005). Carbonaceous chondrites can have lower bulk densities, in the range approximately 1600-2800 (Consolmagno et al 2006) with a lot of sample to sample variation but are less likely to survive bright flight due to the fragility of that meteorite type and will correspondingly be very rare.

The meteorite shape alone is not usually independently accessible via modelling or observation. The shape may also vary during the bright flight phase due to ablation or fragmentation, although it is commonly assumed to be fixed during darkflight. (An exception is the darkflight modelling of (Vinnikov et al. 2016)). One approach, as taken by (Ceplecha 1987) and similarly used for bright flight modelling (Ceplecha et al. 2000; Revelle 2002; Sansom et al. 2015) is to combine these parameters of shape, density, mass to generate a shape-density parameter (since none of these parameters can be independently constrained from observations), and then use values derived from bright flight as an input to darkflight modelling. This approach in theory forces a smooth transition from bright to dark flight, and in practice becomes an issue of observational errors close to the end of bright flight, which strongly influence the choice of values. The fall position prediction is critically related to the choice of drag coefficient (or shape-density factor) of the meteorite; values too low/high will produce over/undershoot in the fall site prediction. This is particularly a contributor very early in the darkflight, when the meteorite is super- and trans-sonic, and is likely travelling non-vertically, as small changes in the choice of drag coefficient here result in large horizontal shifts in the predicted fall position. Once the meteorite is falling terminally, drag coefficient influences the velocity, and hence the time to fall, and controls the influence of winds. Drag coefficient is obviously a function of body shape, and this is usually handled as a scalar factor, by giving the meteorite the drag properties of sphere as function of conditions, and then a multiplier to a non-spherical shape (Ceplecha 1987; Zhdan et al. 2007). Shape of course also factors into angle of attack, which will give variation in drag coefficient, and can even produce lift, however the particular orientation of a meteorite in flight is unknown *a priori* so no aerodynamic lift is assumed. Fragmentation can also cause deviation from the fall line by virtue of the fragmentation event adding transverse velocity to the object, deviating the trajectory (for example the Morávka meteorite (Borovička and Kalenda 2003)).

The details of the drag coefficient behaviours chosen do not appear to be discussed in detail in darkflight papers, with the notable exception of detailed modelling such as (Vinnikov et al. 2016). Apart from meteorite properties, drag coefficient is dependent on many conditions; the velocity, the Mach number, the density of air, all of which vary in dark flight as one goes from the supersonic regime in low density air to a subsonic regime in high density turbulent air, close to ground.

The DFN choice of drag coefficient is detailed in Table 1, extending the earlier table in (Sansom et al. 2015) which focused more on bright flight parameters. Values are generated in separate regimes initially: free molecular flow, and continuum regime. For darkflight conditions, the regime is almost always continuum, which is further divided into hyper/super-sonic, trans-sonic, and sub-sonic. Darkflight terminal falling is subsonic, which is further divided into turbulent and laminar conditions. The choice of regime is parameterised by Knudsen and Reynolds numbers. (The Knudsen number represents the ratio of molecular mean free path distance to a characteristic dimension.) In all cases related to darkflight, the Knudsen number indicates that calculations are in the continuum regime rather than the free molecular flow regime. The Reynolds number is relatively easy to calculate using the standard formulation, taking the characteristic length as the diameter of meteorite, and when compared to the Mach number, one can choose the appropriate regime to estimate a drag coefficient.

As mentioned, the choice of value of the drag coefficient is critical in the early part of the dark flight, however fortunately for meteorite recovery the hyper and supersonic values of drag are estimated to be relatively simple and only slowly changing over a variety of conditions.

In reality, the choice of drag coefficient is probably an approximation to a complex aerodynamic problem. In darkflight, the meteorite may be tumbling, and ablating slightly early on, and complex shapes can generate lift or transverse forces shifting the trajectory. *Post hoc* estimation of the instantaneous drag coefficients and aerodynamic behaviours using a recovered meteorite shape and appropriate aerodynamic modelling software would make an interesting but complex study, not been done to our knowledge.

| Regime | Value | Symbols and Reference |
|---|---|---|
| Free molecular flow Kn > 10 | $$C_{d\_free} = 2 + \frac{\sqrt{1.2}}{2V}\left[1 + \frac{V^2}{16} + 30\right]$$ | Kn = Knudsen Number <br><br> V = velocity, kms-1 <br><br><br> (Khanukaeva 2003) equation 3 |
| Hypersonic | $C_{d\_hyper}$ = 0.92 | (For a sphere) <br><br> (Masson et al. 1960) |
| Transitional from molecular. flow to continuum 0.01< Kn <10 | Bridging function: <br> $$C_{d\_trans} = C_{d\_sub} + (C_{d\_free} - C_{d\_sub})e^{(-0.001*Re^2)}$$ | Re = Reynolds number <br><br> (Khanukaeva 2003) equation 6 |
| Within continuum regime 0.3< M <2.0 | Numerical interpolation using Re and Mach number to graphs and tables in (Miller and Bailey 1979); "Sphere drag at Mach numbers from 0.3 to 2.0 at Reynolds numbers approaching $10^7$", normalised to match the boundary conditions of $C_{d\_trans}$ and $C_{d\_sub}$ | |

| Where subsonic drag coefficient is | $C_{d\_sub} = \dfrac{24}{Re}\left[1 + exp(2.3288 - 6.4581\varphi + 2.4486\varphi^2)Re^{(0.0964+0.5565\varphi)}\right]$ $+ \dfrac{Re * exp(4.905 - 13.8944\varphi + 18.4222\varphi^2 - 10.2599\varphi^3)}{Re + exp(1.4681 + 12.2584\varphi - 20.7322\varphi^2 + 15.8855\varphi^3)}$ | $\varphi$ = sphericity (Haider and Levenspiel 1989) equation 11 |
|---|---|---|

*Table 1 - detailing the regimes and derivations of drag coefficients used in darkflight calculations. We have implemented this as a callable function in Python 3, available at www.github.com/desertfireballnetwork/DFN_darkflight.*

## 2.3 Atmospheric Wind data

During the descent through the stratosphere and troposphere, the atmosphere is not quiescent. Upper atmosphere winds and density variations can deflect the falling meteoroid, such that the ground impact positions may be shifted by several kilometres. In particular, upper atmosphere phenomena such as jet streams are the major drivers of how the fall line is shifted relative to an analysis without considering winds. To predict these atmospheric properties, the DFN uses the NCAR atmospheric modelling system Weather Research and Forecasting (WRF) version 4, with ARW dynamic core. (Skamarock et al. 2019)[3]. The WRF is a forecast model that incorporates real world data (such as balloon flights) to model atmosphere dynamics, capable for being initialised from a global data set to generate mesoscale results at high spatial resolutions suitable for inputs into a darkflight calculation. The WRF software generates a weather simulation product as a three-dimensional data matrix in a latitude/longitude/height cuboid around the bright flight end point. From the model, grid values are extracted for the atmospheric properties of relevance to darkflight. This includes the pressure, density, temperature, relative humidity, horizontal wind speeds as function of height, latitude, and longitude (in u, v coordinates), such that these can be interpolated during darkflight modelling in 3D to precise locations. Since the WRF data cuboid is not necessarily north-south orientated, wind values must be extracted; this extraction therefore involves a coordinate transform, as WRF grids are not typically aligned with true north, and local verticals may also need to be corrected. For convention in our calculations, we define wind directions in degrees, with North=0, East positive, with a positive wind magnitude in direction of wind travel, not the wind's origin.

The primary use of the WRF tool is weather forecasting. The top level forecast is done on a global matrix - extrapolating the state of the weather matrix, based on past observations using a physical model of the atmosphere, to the future. The global matrix is typically applied with a time step/resolution of 6 hours. This top level product then initialises finer resolution modelling over a smaller area in order to achieve better precision both in time and space. To get detailed (fine grid) forecast for a local area, a set of embedded domains is defined (typically 4 levels), increasing in cell size resolution, to finally achieve typically 1 km resolution around the bright flight end point. Each domain is based on the physical model of the atmosphere, with the boundary states coming from higher level matrix points. However, in the darkflight modelling case, we are not forecasting, but interpolating the past state of weather. We do this along the meteoroid dark flight trajectory using a physical model of the atmosphere based on the observations rather than forecast, using the archived data from the NCEP FNL Operational Model Global Tropospheric Analysis online datasets[4]. These archived snapshots contain constraints on the global weather conditions at each time step, with a six hour interval. When hindcasting the conditions, one starts with a snapshot, and propagates the weather model forward in time (and at higher spatial resolution in the location of interest). The propagation has forcing conditions, such that the results generated must pass through the conditions recorded by later snapshots, including snapshots of times after the meteorite fall.

---

[3] https://rda.ucar.edu/datasets/ds083.2/

Due to the stochastic nature of the WRF numerical modelling software, slightly different results are produced each time it is run, even with the same input data, but the model outputs do not provide any error analysis. Variations arise from floating point precision, and from the limitations of the hardware and numerical libraries used. As its primary purpose is weather forecasting, the success of the modelling is evaluated by comparing the forecast with real weather observations. It is also worth noting that observational data for central Australia is somewhat sparse. To resolve this lack of defined error bars, for each studied case of possible meteorite fall we run models starting from different archived global snapshots, typically 0-6 hours, 6-12 hours and 12-18 hours before the fall time, and then extract from the results the conditions and wind profile at the time of the meteorite fall. Comparison of these multiple cases highlights how stable the weather was, to explore uncertainties in the product introduced by the known errors in the observational data and the ability of the model to work for specific weather situations. For example, a stable weather situation is more likely to give very similar sets of wind profiles, while a cold front passing shortly before the time of the fall can be expected to produce a lot more diversity in the vertical profile plots extracted from the three individual modelling products of the different time windows. The amount of profile variation provides insights as to how to constrain our Monte Carlo darkflight simulations (as described below).

## 2.4 Implementation details

For the DFN operations, we have implemented a darkflight integrator in Python 3, using the SciPy, NumPy and AstroPy libraries as needed (Jones et al. 2001; Astropy Collaboration et al. 2018). SciPy includes the core Runge-Kutta integrator function. We use the WRF-python library provided by NCAR to access the data files produced by WRF, and several smaller libraries are used for geocentric coordinate transforms. The resulting tool takes as an input the results from a triangulation, plus putative meteorite characteristics, and a data file from a

71   WRF scenario, and produces time-position trajectories to ground. This tool can then be
72   iterated over to investigate multiple scenarios, to produce likely fall positions for practical use
73   for ground meteorite searching.

74   This darkflight integration can be carried out in an Earth inertial coordinate system, or in an
75   Earth Centre Fixed frame (where Coriolis force must be included as the atmosphere is coupled
76   to the Earth's rotation on short timescales), A geodetic Earth model is used, and the gravity
77   vector is calculated as perpendicular to the Earth's reference ellipsoid.

78   The resulting product of darkflight aims to predict a likely search area for a meteorite. The
79   most basic result is a fall line - a ground plot showing a line giving fall positions for a given
80   range of proposed masses. For the DFN, a wide range of masses are modelled, to aid in
81   searching planning; generally this range is much larger than the expected uncertainty of the
82   final mass, which is obtained from bright flight modelling (Sansom et al. 2020). This can be
83   produced for multiple scenarios such as different assumed shapes or wind profiles, resulting
84   in multiple fall lines. In Figure 2 we plot this simple case for the Murrili fall, discussed in more
85   detail in following section. We show two scenarios: idealised spherical and a brick-shaped
86   meteorites. Note the curved shapes of the fall line, and the offset caused by shape choice; the
87   curve is a result of the influence of the atmospheric winds, whereby drag coefficients are
88   generally higher for smaller bodies.

89

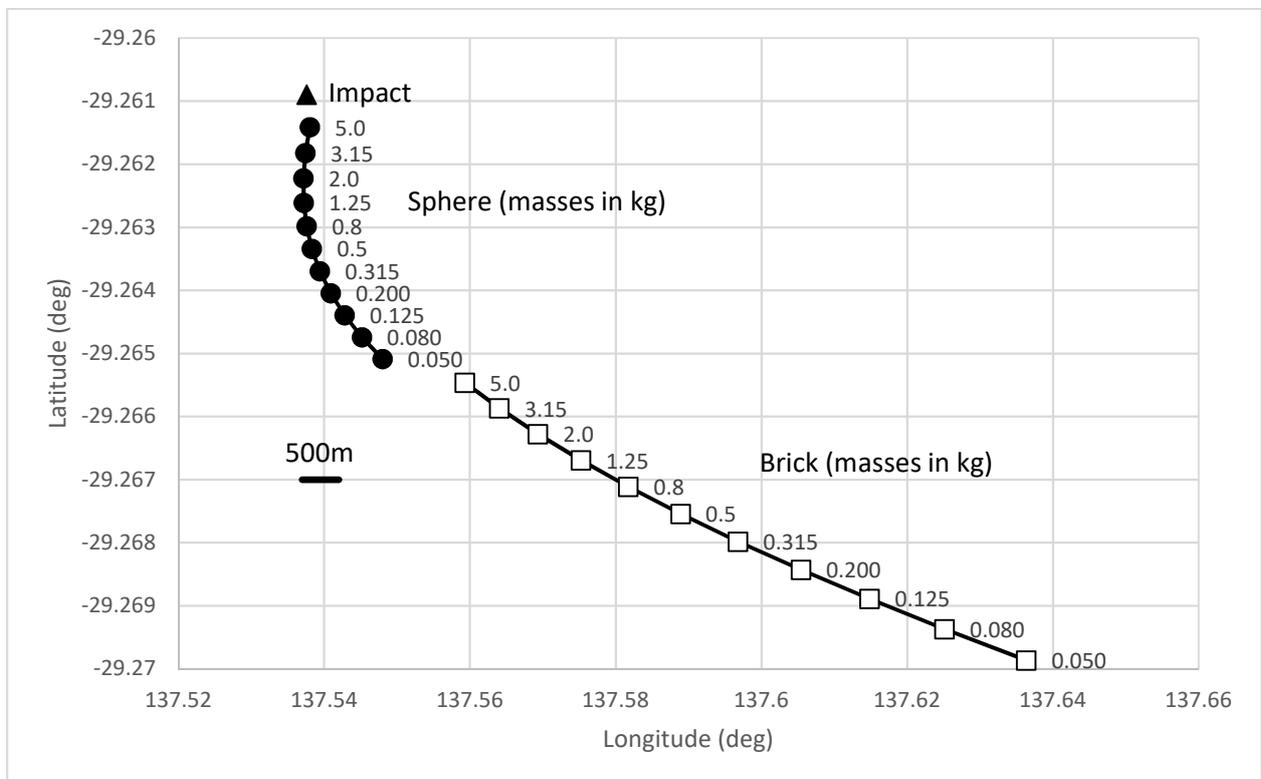

90

 *Figure 2 - Fall line predictions from the Murrili bright flight endpoint, assuming a chondritic sphere, and brick, using*
*the wind model 06:00 snapshot as described in the following section. The actual recovered fall position is marked*
*with black triangle.*

For searching, of greater use is an impact probability scatter plot or heat map. To generate
heat maps, a Monte Carlo approach is used, varying both the meteorite input parameters, and
the atmospheric variables. Monte Carlo modelling can be computationally expensive, but as
a bare minimum, several scenarios can be calculated since the meteorite shape/density/mass
is estimated from modelling (Sansom et al. 2017) but not known with certainty. The dimensions
of the Monte Carlo scatter can be used to inform the likely width of the uncertainty in the fall
line, which allows searchers to prioritise their activities in the most time-efficient manner. To
highlight all the subtleties and complexity involved, we describe in detail the analysis related
to the Murrili meteorite fall (Sansom et al. 2020).

## 3 MURRILI METEORITE FALL AS A CASE STUDY, WITH DISCUSSION

The Murrili fireball provides an interesting case study of the effects and importance of detailed
modelling of the darkflight. The fireball that resulted in the Murrili meteorite occurred over Kati-
Thanda (Lake Eyre south), in South Australia at 2015-11-27T10:43:45.5 UTC. The fireball was
observed by the DFN, and the meteorite recovered within the following month. Hence the
recovered shape, density, and mass can be used to back validate darkflight modelling. Murrili
is a fortunate case as the scenario was quite ideal from the point of meteorite recovery
(although the ground conditions were difficult): The fireball had a well observed bright flight,
almost equidistant between four DFN all sky cameras, that all captured the full event with high
quality data (Wilpoorina, William Creek, Nilpena and Etadunna), supported by two more
distant cameras at Billa Kalina and Mount Barry that also contributed. The trajectory was
relatively steep, with a zenith angle of 21.8$^O$, and the fireball penetrated deep into the
atmosphere, to a low altitude of 18.0 km. The triangulation, modelling and recovery is
described in detail in (Sansom et al. 2020).

The steepness, and the low final altitude dramatically reduce the influence—and hence the
associated uncertainties—of the local wind conditions, reducing the errors compared to less
favourable examples, such as a shallow entry angle with an end point altitudes that could be
much higher (over 30 km in some cases). This combination of factors makes Murrili a case
study where one can investigate the limitations of darkflight modelling – the almost ideal
experimental situation should result in predictions that closely match the recovered fall
position, provided the model is accurate.

125 Further confidence for meteorite searching and fall position was given by aerial
126 reconnaissance of the site, using a light aircraft from William Creek (by authors MC and BD),
127 that observed a visible splash in the lakebed, at the area of the expected the meteorite fall.
128 Approximate coordinates from the light aircraft were used to begin ground searching. This
129 single splash also supported the lack of significant fragmentation seen in the bright flight
130 images.

## 3.1 Constraints from bright flight observations

132 The factors as mentioned above resulted in a low uncertainty in the end position and velocity,
133 as detailed in Table 2.

134

| Date/time | 2015-11-27T10:43:51.626 +/- 0.05 |
|---|---|
| Longitude, degrees East positive | 137.47817  +/- 50 m |
| Latitude, degrees | -29.26534 +/- 50 m |
| Height above WGS84, m | 17960 +/- 40 |
| Velocity, m/s | 3280 +/- 210 |
| Zenith angle, degrees (from vertical) | 21.80 +/- 0.05 |
| Azimuth angle, degrees (north = 0, clockwise positive) | 82.60 +/- 0.05 |

135 *Table 2 showing the end of bright-flight parameters, used for the initiation of darkflight calculations. See (Sansom*
136 *et al. 2020) for further details.*

137

138 As discussed in detail in (Sansom et al. 2020), the final mass prediction was for a value of
139 1.9 kg, +/- 0.4 kg, assuming a chondritic density of 3500 kg/m$^3$. The best fit modelling to the
140 luminous trajectory using first the α-β approach of (Gritsevich 2007; Sansom et al. 2019), and
141 then an Extended Kalman Filter (Sansom et al 2015) was relatively smooth, showing no
142 evidence of major fragmentation, and giving a Shape Change Coefficient which matches well
143 with typical values, so this event does not stand out significantly as indicating an unusual
144 shape, or any strong deviation from expected path (such as might be from lift or strong
145 asymmetry.

## Single object (sphere) darkflight integrations

147 We begin by considering a integration case, and discuss the effects of factors such as wind
148 modelling in the following sections. In Figure 3 we plot the calculated sphere drag coefficient
149 ($C_d$), and fall velocity as function of altitude. We see that $C_d$ does change significantly during
150 descent, as velocity and atmospheric density both vary, in particular early on when velocity is
151 still dominated by the arrival velocity, rather than terminal effects. Darkflight encompasses the

152   full supersonic to subsonic regime, and this illustrates the importance of varying $C_d$. Murrili is
153   almost an ideal case, with short darkflight, this effect would be more pronounced for a
154   meteorite at say a shallow angle, from a higher altitude.
155

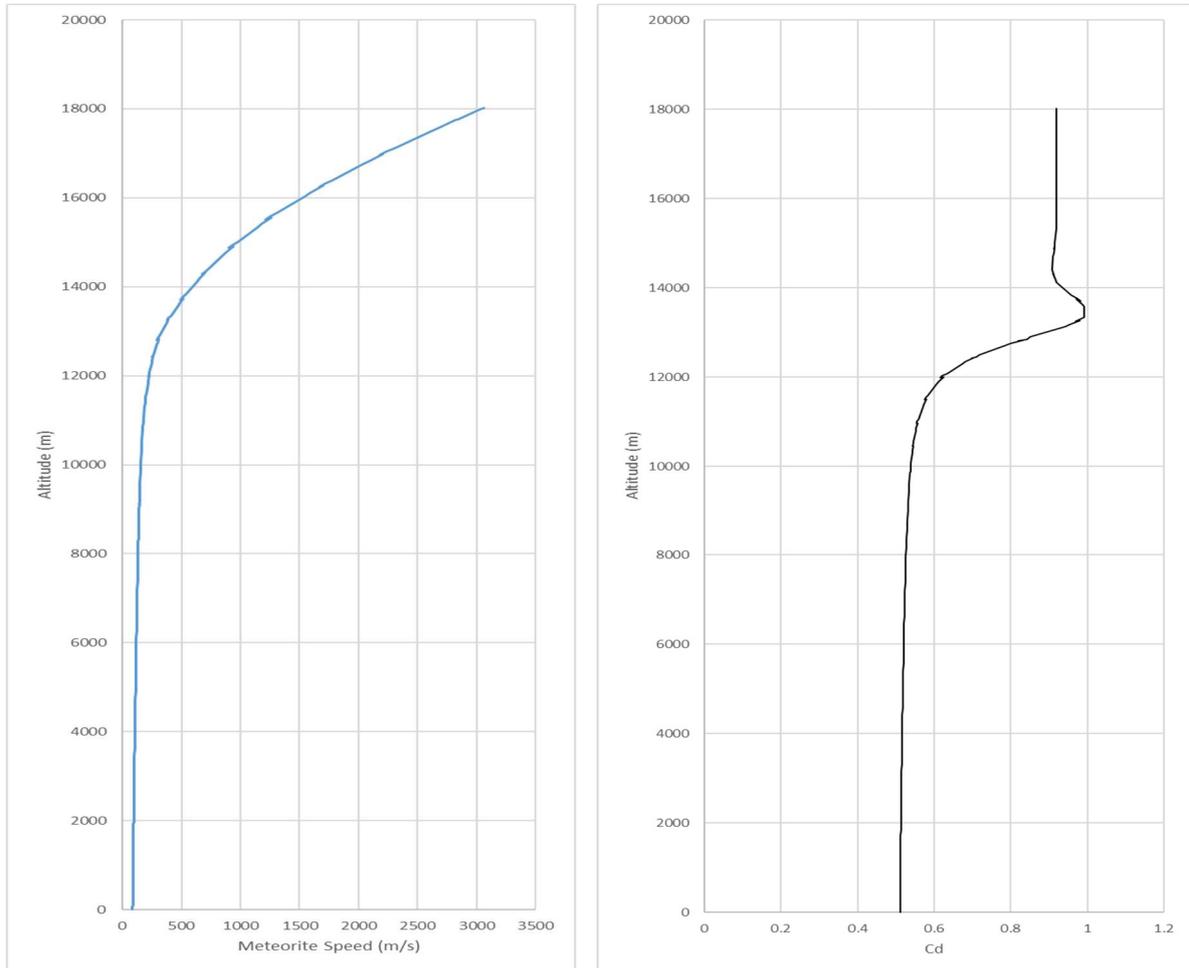

156   *Figure 3, showing modelled Murrili fall speed, and drag coefficient as a function of altitude*

157   In Figure 4 we show the idealised model trajectories viewed from above for several chondritic
158   spherical masses, equivalent to generating the sphere fall line shown in Figure 2. This
159   illustrates the effect of winds on fall position, and how the fall line is constructed. Even in the
160   case of Murrili, with a particularly low end point and steep trajectory, a 1kg sphere is still
161   deviated significantly in predicted fall position compared to a no-wind scenario.

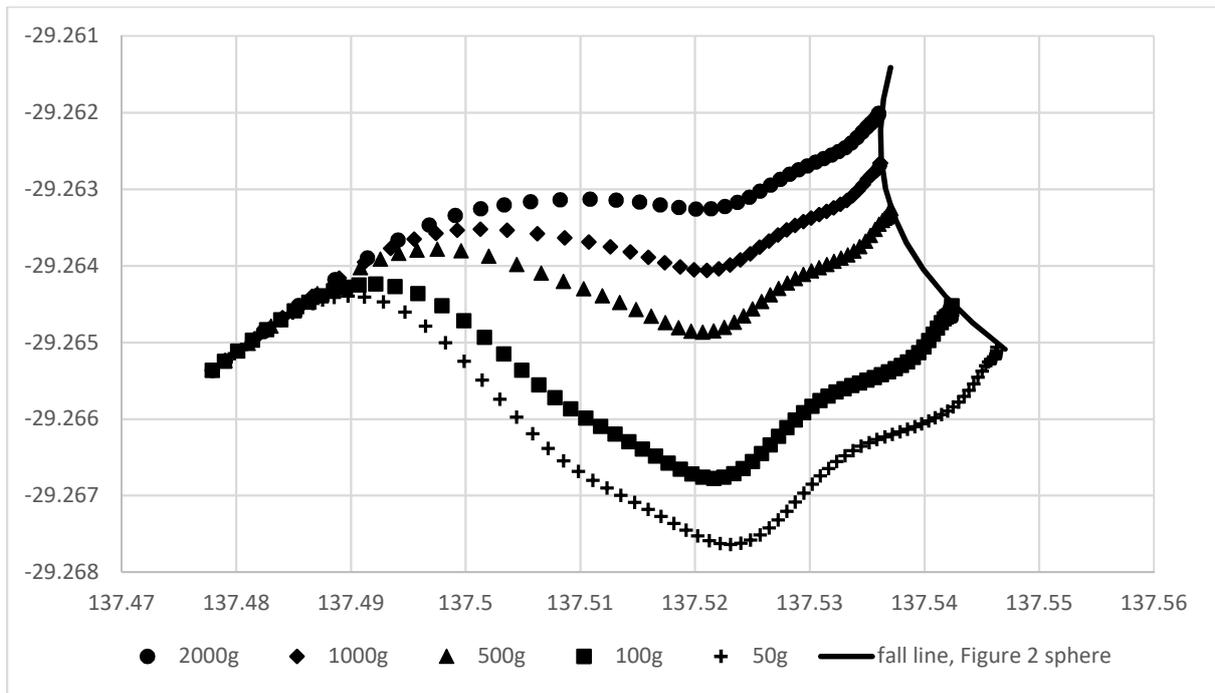

162

*Figure 4 modelled trajectories of chondritic spheres for different masses in longitude and latitude, showing the*
*influence of winds and atmosphere during descent, from Murrili bright flight endpoint.*

## 3.2 Effects of variation in wind profiles on fall lines.

Wind data is generated using the WRF model. In the case of the Murrili meteorite, we used
version v3.7.1 for the fall coordinate predictions prior to meteorite searching, and later v3.9.1
as it came available for re-running of the initial analyses (Skamarock et al. 2008).

As is seen in Figure 2 and Figure 4 , the atmospheric winds distort and shift the fall line, in a
mass dependent manner. However, this wind profile used is a modelling prediction generated
by WRF, with no way to be independently directly verified at this locale.

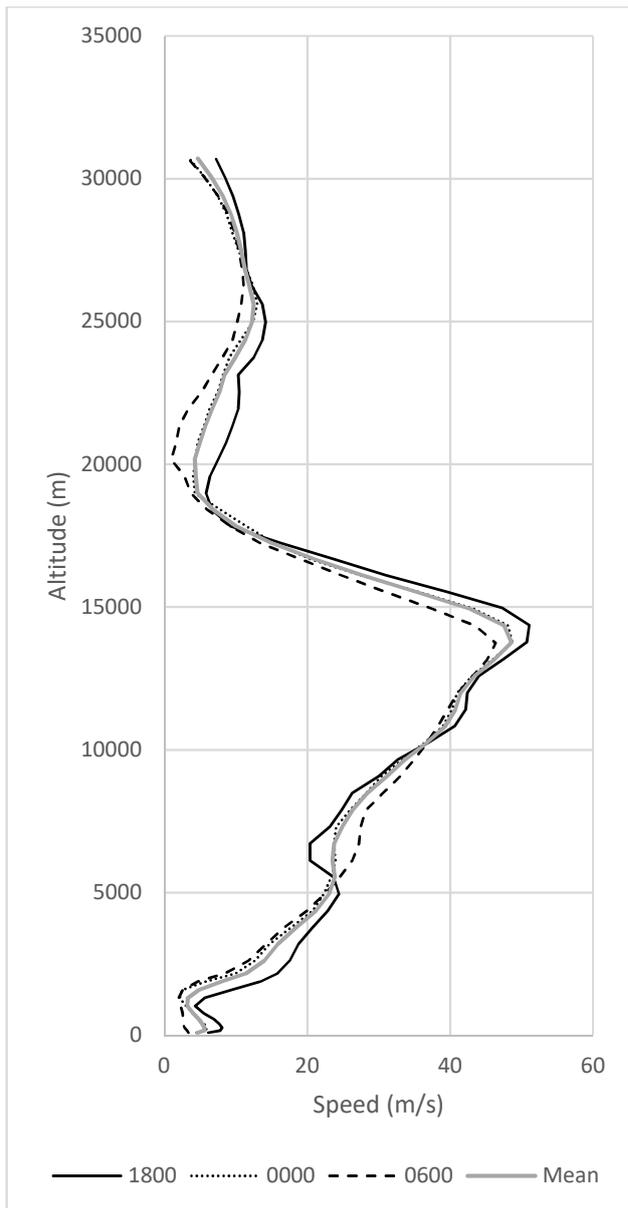 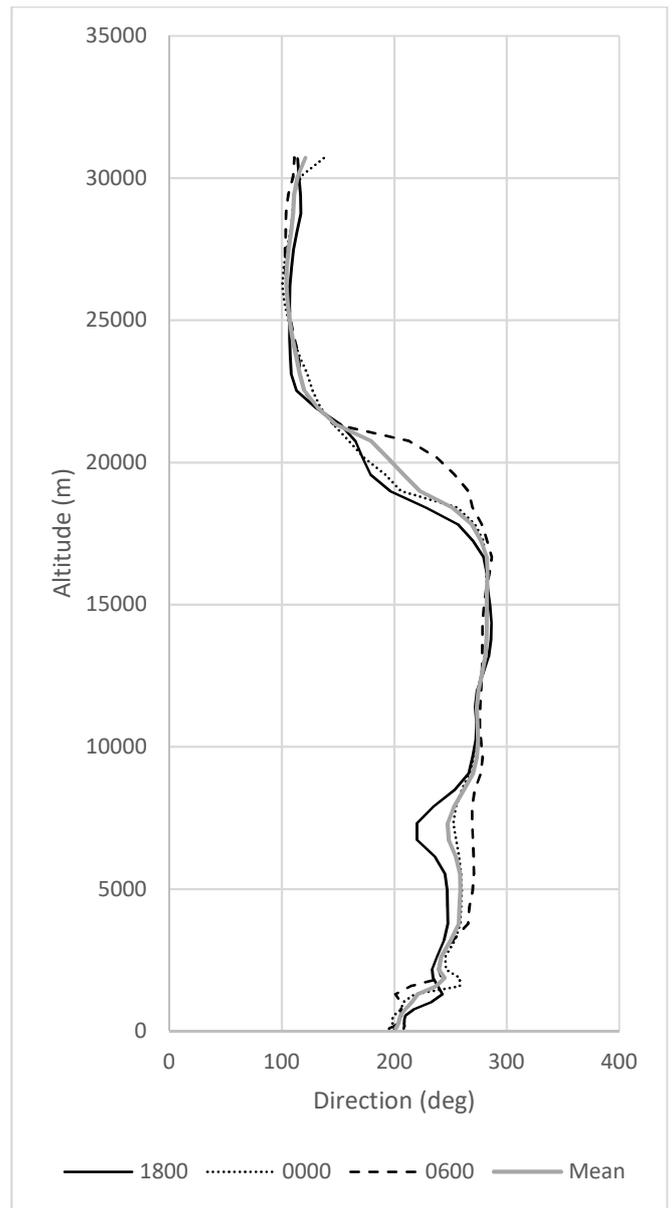

*Figure 5 plot of wind profiles from each WRF model run for comparison. Dominating westerly winds at altitudes 10-15km are typical for the subtropical jet stream in the area of the fall. The individual profiles are products of WRF runs starting from the snapshot 2015-11-26 18:00 UTC, 2015-11-27 00:00 UTC and 2015-11-27 06:00 UTC.*

As mentioned, WRF can be initiated with archived global snapshots at six hourly intervals. Examination of the spread of profiles from these snapshots is one method to generate an estimate of the plausible variation in wind models to use within any Monte Carlo simulation. In Figure 5 plot of wind profiles from each WRF model run for comparison. Dominating westerly winds at altitudes 10-15km are typical for the subtropical jet stream in the area of the fall. We plot results from three model runs based on different snapshots. Note that the winds are exceptionally strong at the 15,000 m levels, indicating a jet stream effect, and in fact greater

185 than have been observed in most other cases investigated by DFN. There is also some
186 variation between outputs from each snapshot. Regional historical weather maps for this area
187 of Australia for the 25-30[th] November 2015 show a high pressure region passing to the south
188 of the fall area, with a change in general wind direction on the 26-28[th] November. The precise
189 timing of this change may well have taxed the fidelity of WRF to carry out a high spatial-
190 temporal resolution simulation far from actual archived weather observations.

191 One can investigate the gross effect of this wind variation by carrying out darkflight calculation
192 for Murrili for ordinary chondritic-density spheres using the results from each wind model, as
193 shown in Figure 6.

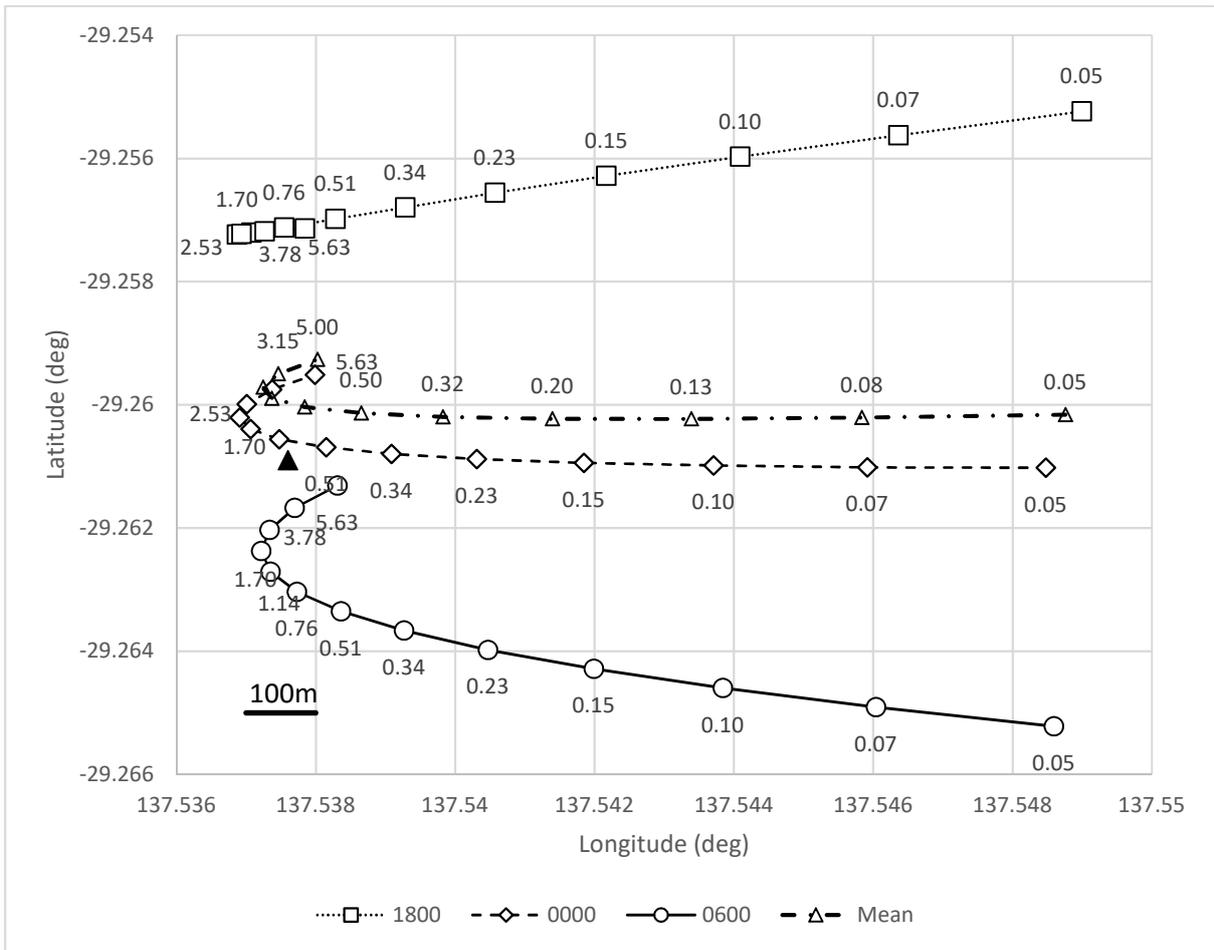

194

195 *Figure 6 Variation in fall line predictions for chondritic spheres due to variations in WRF wind model data, model*
196 *snapshot shown in legend. Numbers along line indicate the mass on ground. Some numbers omitted for clarity on*
197 *some lines, but the points shown correspond to matching mass markers on the lowermost curve. The final*
198 *recovered meteorite fall position (1.68 kg) is also shown with a solid triangle (▲).*

199

200 There are significant variations between ground predictions from the wind models, typically
201 200-300 m between lines. This strong wind dependence is in part a consequence of the
202 structure of the drag equation, where relative velocity is a squared factor (Equation 1). In the bigger picture of conducting ground searches in the Australian outback, this can be an issue, as accepting such variations produces an unrealistically large search area that is not feasible to search (compounded by the other uncertainties discussed in following sections). Ironically, it appears that although the Murrili triangulation scenario was ideal, darkflight conditions were poor. Of some hope in the case of Murrili is that most plausible mass ranges—from bright flight estimates (Sansom et al. 2020)—are close to the western edge of the fall lines, at the areas of most fall line curvature. So in this case, this means that the area that needed to be searched was relatively constrained, regardless of the choice of wind models.

More generally when wind model scenarios diverge, at one extreme one can elect to search all areas, or one can prioritise. For this and previous searches, when faced with this choice, the DFN has usually elected to focus on the penultimate/2nd-shortest WRF model outputs; in this case starting from the snapshot from 2015-11-27 00:00 UTC. This is purely an empirical choice, based on backwards comparison of almost all of the DFN recovered meteorites (and with hindsight a good match to Murrili as well) (Spurný et al. 2012; Spurny et al. 2012; Sansom et al. 2020). This may be a result of the mechanics within WRF that implies that longer runs are needed for accuracy, to allow the WRF model to achieve numerical stability, whereas the longest simulations allow deviations from reality to accumulate. We lack the expertise in climate modelling to comment in detail, and this is clearly an area that needs further study and more data, as very few meteorites exist with both known precise end points, and well characterised, published, detailed shapes and densities. As such, our approach is to empirically use the penultimate wind model, ensuring that searching in the field is aware of the limitations of this approach. It is worth noting that it appears that Murrili is a particularly variable WRF scenario; in previous DFN searches the fall lines from different WRF snapshots are often closely overlapping, such that it is possible to search all scenarios within a reasonable timeframe and a judgment on choice of wind profile is not required. In this case, it was fortunate that the fall was on a salt lake, allowing the use of quadbikes to search large areas relatively quickly in comparison to foot searching in a vegetated area.

However, wind is not the only uncertainty affecting fall position in darkflight calculations; one must also consider shape and meteorite density, which can be partially constrained but is effectively unknown.

### 3.3 Effects of shape and density

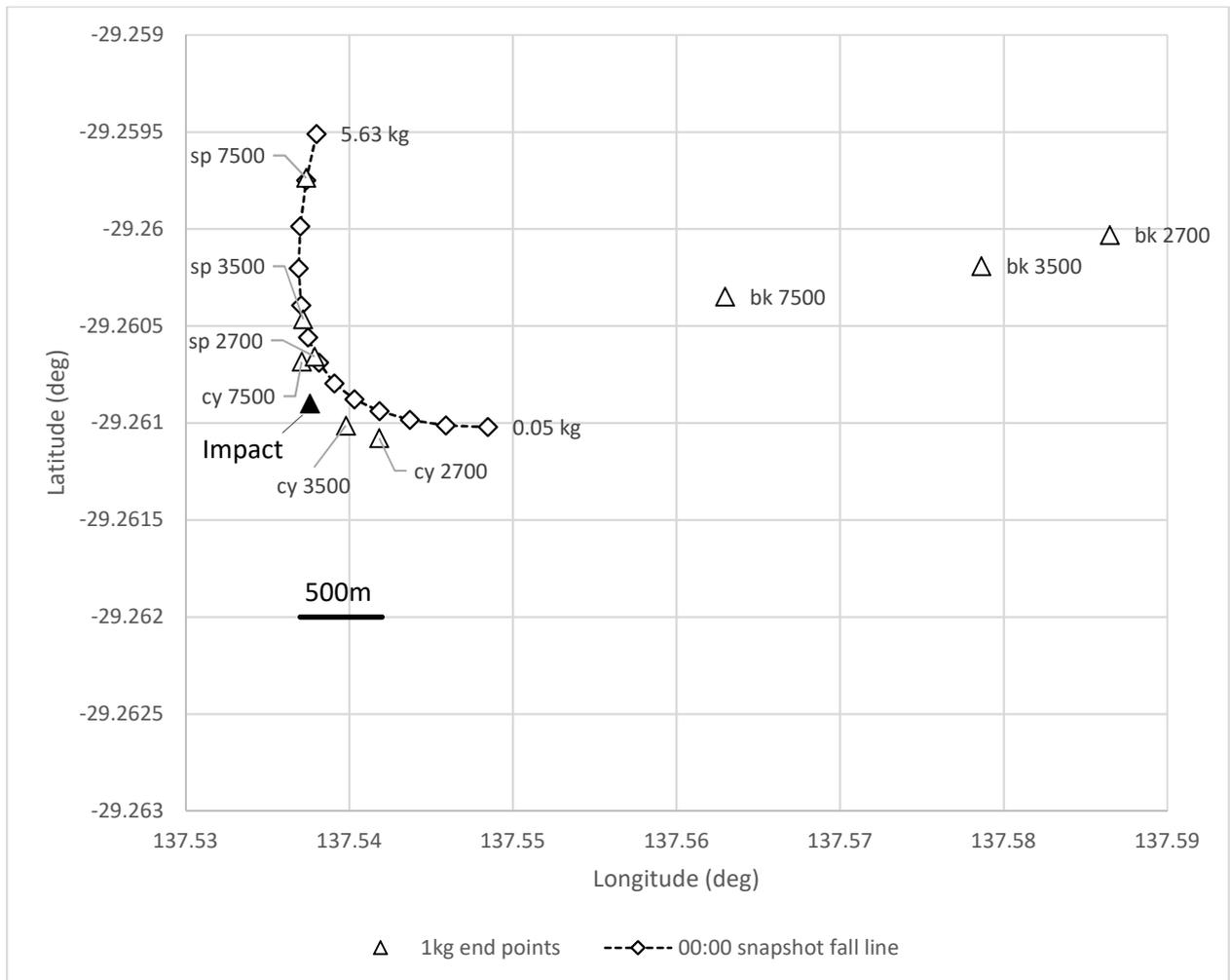

235

*Figure 7 plot showing the effects of density and shape on final position of a hypothetical 1kg object. Shapes plotted*
236
*are sphere, cylinder, and brick (as defined in the text). Density of the object is constrained to 2700, 3500, and*
237
*7500 kgm⁻³. For comparison we include the fall line based on the 2015-11-27 00:00 UTC wind snapshot (3500 kgm⁻*
238
*³ sphere, varying mass) and the impact point from Figure 4*
239

240

To investigate the effect of meteorite shape and density choices on predictions, Figure 7 plots
241
darkflight predictions for three different densities (2700, 3500, 7500 kg/m³), and three different
242
shapes (sphere, cylinder and brick, as defined by (Zhdan et al. 2007), compared to the 00:00
243
snapshot fall line. The change in shape (which effectively changes drag coefficient) and
244
density (which changes cross sectional area for the same mass), have direct effects on the
245
fall line position, and in this case (and in other cases seen by the DFN (Devillepoix et al.
246
2018)), choice of shape has a greater influence than meteorite mass or density prediction. For
247
density variations, this results in same mass falling on effectively the same fall line but
248
translated along the line. This is also the case when changing shape from cylinder to brick,
249
however this shift is more extreme. From the DFN experience of recovered meteorites a 2.5 x
250
1.5 x 1 brick shape will be an outlier; in general, recovered samples seem to be best fit by
251

drag coefficients close to spherical. Furthermore, it is worth noting the dominance of shape choice: effective drag coefficient is changed by a factor of 1 to 2 (Zhdan et al. 2007), and although density effects are varying comparably (through the cross-sectional area), the shape effects dominate. This effective along-line shifting effect has advantages and disadvantages for searching. Traverse distance from the line that must be searched is essentially unchanged, but more of the line must be searched, given a particular mass range prediction, as shape will shift this further along the line. However, since multiple scenarios overlap, several can be effectively searched at the same time.

### 3.4 Monte Carlo studies of fall line variations and scatter

To combine all these factors, and generate some understanding of the associated uncertainties that predictions would generate in the case of Murrili, we have carried out Monte Carlo simulations of darkflight descents covering ranges

- with mass in range of 1.5-2.3 kg, based on pre-recovery predictions from (Sansom et al. 2020)
- with a density 3500 kg/m$^3$
- first with no atmospheric winds, then allowing variation of up to +/-5 % in wind magnitude and direction for a wind profile, using the data based on the 00:00 snapshot, chosen for reasons discussed in section 3.2. The 5% uncertainty is derived from the deviation of the profile variations seen in Figure 5.
- meteorite shape can vary from spherical to a rounded brick shape—defined as 2.5 x 1.5 x 1 brick dimensions with corners smoothed off, using the rationale and estimates of (Zhdan et al. 2007), with the highest drag direction of the brick orientated in the direction of travel. For wind effects acting transverse to the orientated brick, we use the drag coefficients from (Zhdan et al. 2007) for a cube shaped object. (For non-spherical objects in supersonic regimes, the object usually settles into an orientation with the maximum cross section across the trajectory, giving the maximum drag coefficient, as detailed in (Turchak and Gritsevich 2014) and references therein.)

These Montecarlo ground positions are then displayed as scatter maps in Figure 8.

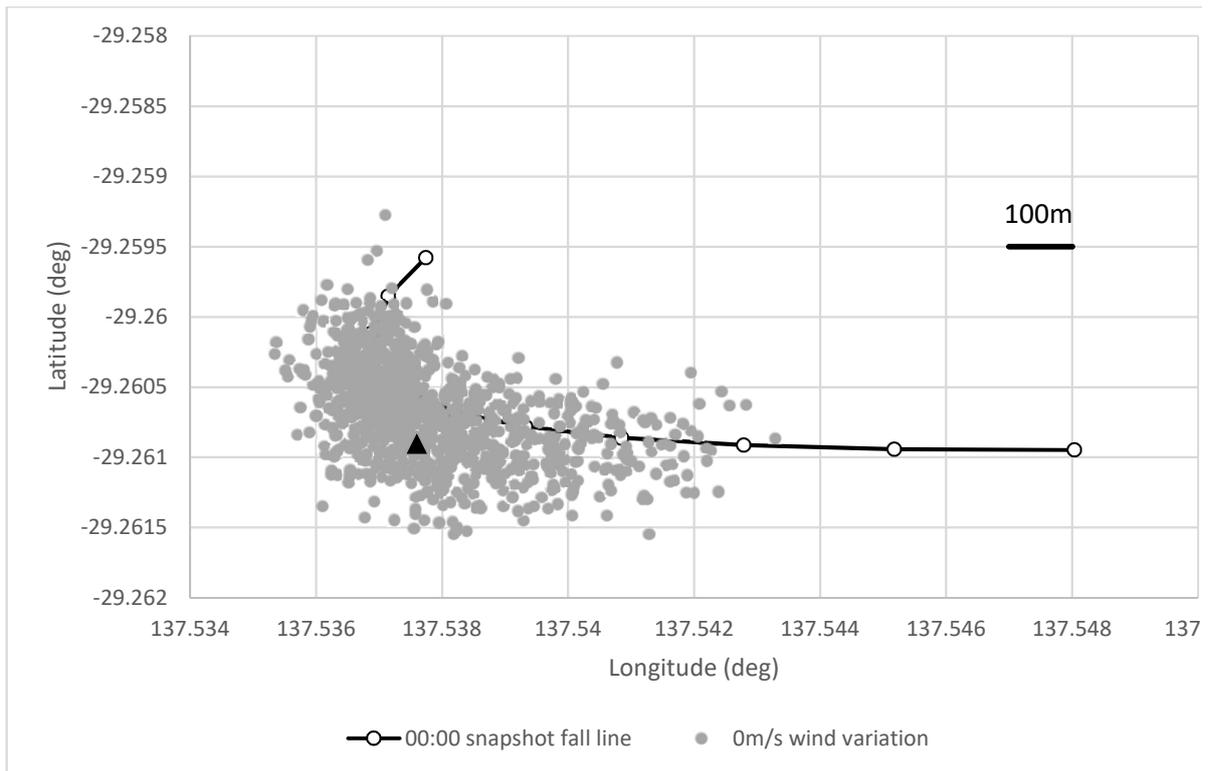

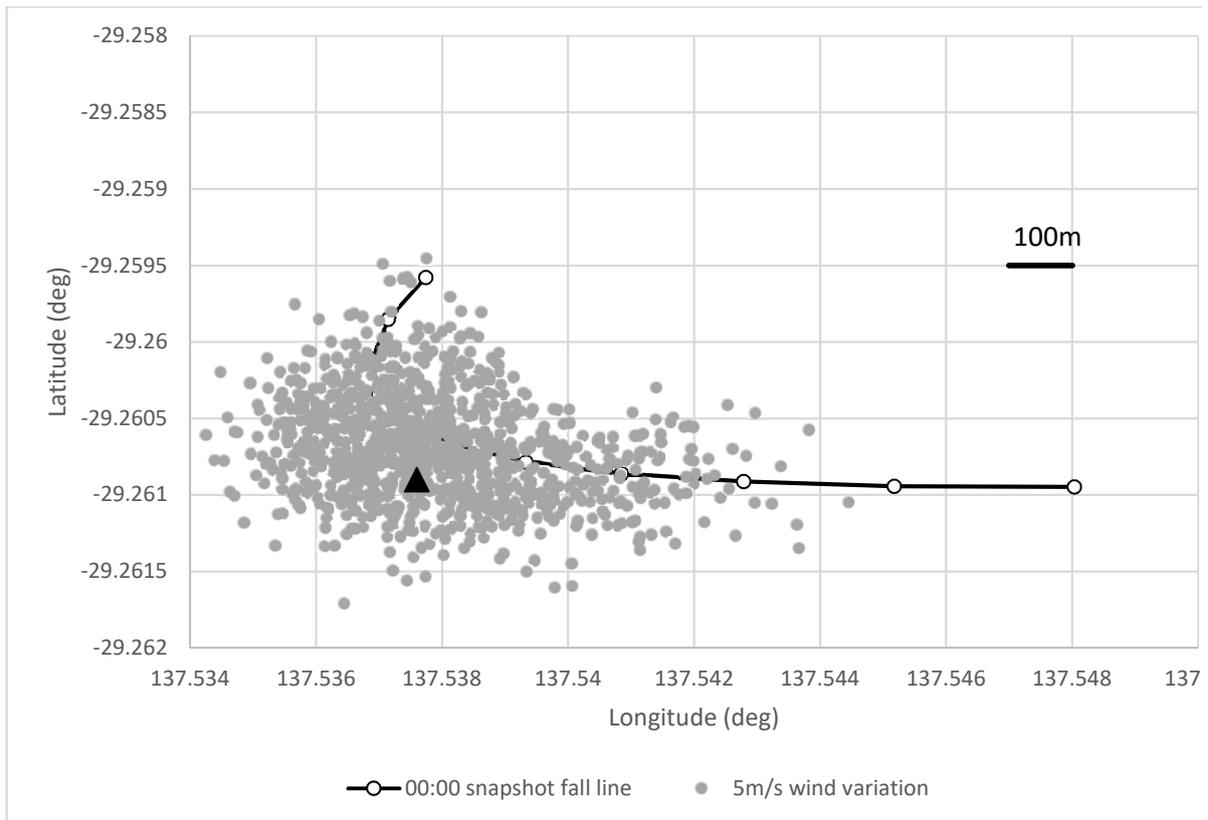

Figure 8 (a) Monte Carlo results of 0.5-2kg chondritic meteorite, 1000 runs, allowing initial shape to vary from sphere to rounded brick (b) same but with wind variation of 5% allowed in direction and magnitude of the 00:00 snapshot based wind profile provided by WRF. The simulations also allow variation in the initial vector from end of bright flight triangulation, using the uncertainties in Table 2.

The ground scatter plots are very roughly the same location, although with greater scatter as wind due to wind. In overall dimensions, the scatter distributions are about 200 m orthogonal to the fall line, indicating a reasonable distance to search from the fall lines, and rough length along fall line of both scatters is 400 m, giving each a searchable area of about 0.2 km².

## 3.5 Fall line prediction compared to meteorite recovered position

The previous sections describe the analysis that can be done before meteorite recovery, using shape approximations and wind model predictions. We now consider after the meteorite recovery, when the actual shape, mass, density are known, and a newer version of the WRF is available. Despite the high quality of triangulation and the low end point of the bright flight, the fall position was ~40 m away from the preferred line prediction, and ~100 m along the line from a sphere-based prediction. This would at first glance appear excellent from a practical searching point of view, but for a less favourable fall with a higher end point this offset would be proportionally larger. For a shallower entry angle, fall line uncertainties also increase due to greater horizontal travel at high velocity immediately after the end of bright flight, where any unknowns in the drag coefficient and shape contribute greatly. One should then investigate possible causes for this orthogonal offset: Since many factors are constrained by the properties of the meteorite, one is essentially left with issues of wind models accuracy, a non-ideal shape, or modelling issues such as choice of drag coefficient. The preceding analysis has focused on the data available prior to recovery, but for the following figures using post recovery data, fall lines are plotted use a later recalculation of the wind models, using WRF v3.9.1, which has shifted the fall line predictions slightly, by about 100 m to the south.

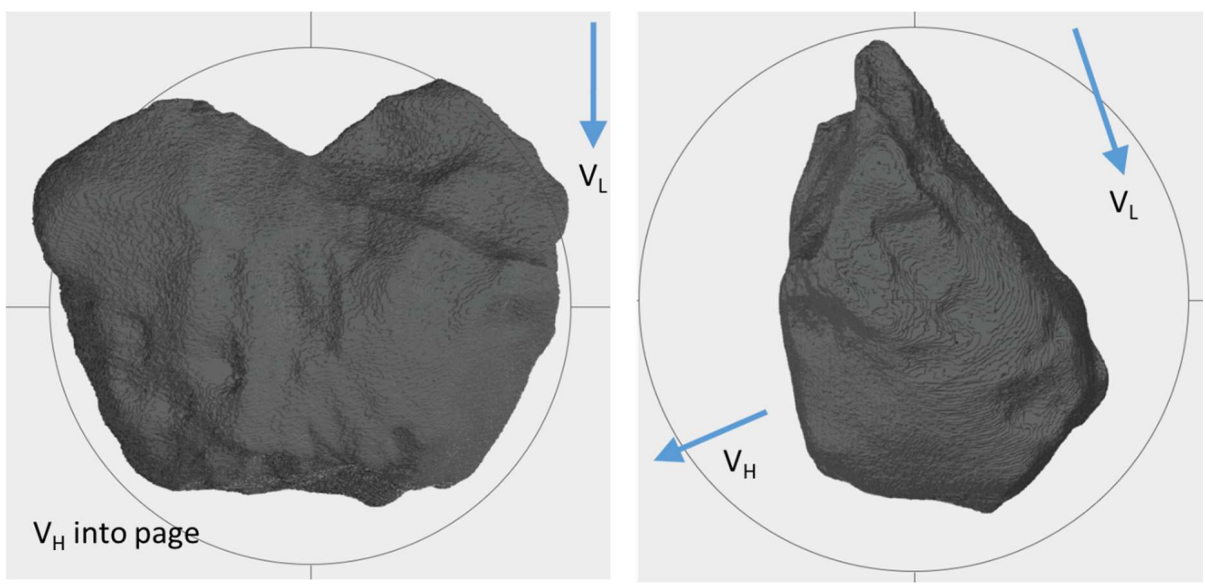

*Figure 9 Murrili meteorite dimensions and shape. Compared to a sphere, the rock is flattened and heart-shaped, but almost teardrop from the side view*

311

312   The Murrili meteorite is shown in Figure 9. Its extents in (a) are approximately 130mm x 90mm,

313   while the thickness in (b) is 70 mm. We round to the nearest 5 mm due to fine detail irregular

314   variations in surface features. The meteorite volume (obtained from a CT scan) is

315   474,731 mm$^3$, giving an equivalent sphere radius of 48 mm (volumetric radius). Alternatively,

316   the meteorite surface area is 40,299 mm$^2$, giving a sphericity of 0.71 or 0.73 depending on the

317   definition used; surface area of an equivalent volume sphere over surface area of meteorite is

318   28,952/40,299 giving 0.71 (Pettijohn 1975), or alternatively equivalent volume sphere

319   diameter over diameter of circumscribing sphere is 96 mm/130 mm giving 0.73 (Wadell 1935).

320   As part of the meteorite description studies prior to official classification, the meteorite was

321   visually inspected, and we have reviewed the 3D CT scan data to investigate any fusion crust

322   features that might hint at orientation or even changes in flow regime during descent – there

323   are some possible flow lines, but they are very faint and quite subjective, and not completely

324   compelling. Unfortunately, the meteorite fell into a wet salt lake in the Australian summer, and

325   so was buried for about a month in warm saltwater mud before recovery. The extensive

326   weathering appears to have removed a lot of fine detail, preventing any firm conclusions about

327   orientation. We have also reviewed the details of the impact site for insight into orientation, via

328   images of the impact itself, and consideration of the impact velocity. The impact appears

329   roughly circular in form (Sansom et al. 2020), although the salt lake crust may break unevenly.

330   (Sansom et al. 2020) additionally states that meteorite was buried 42 cm into the mud.

331   However, the lack of knowledge of the properties of the ground, as needed for the appropriate

332   projectile depth penetration equations (Young 1967, 1997) can generate a wide range of

333   velocities, providing little guidance on reconstructing the impact velocity.

334   One can attempt to correct the spherical drag coefficient with some factor based on the known

335   shape of the recovered meteorite. For non-spherical bodies, this problem has been studied in

336   the context of dust settling rates, often in relation to industrial processes or environmental

337   studies. See for example (Connolly et al. 2020) and references therein, or (Kleinstreuer and

338   Feng 2013) for a review from a biomedical context. For settling rate studies, the Corey Shape

339   Factor, CSF, ($d_{min}/\sqrt{(d_{max}*d_{med})}$) where d is diameter, (Corey 1949), is the most commonly used

340   approximation and provides the most data for correlations between publications. For Murrili

341   this evaluates to 0.64. However, one must exercise caution with using CSF outside of settling

342   studies – a CSF of 1.0 describes a sphere, a cube or several other regular solids, which all

343   have different drag coefficients.

344   Free fall drag coefficient has also been derived as a function of sphericity from empirical and

345   theoretical studies ((Haider and Levenspiel 1989) and references therein). Within the DFN

346   general darkflight code implementation it is possible to explicitly specify sphericity (Table 1,

347   forcing the use of Haider and Levenspiel, equation 11), overriding the default calculations for

348  a sphere. (Hölzer and Sommerfeld 2008) extend this formulation to include projectile
349  orientation, by treating crosswise and longitudinal sphericity separately. To investigate and
350  compare this, we have implemented their equation 9, and then calculated Monte Carlo
351  darkflight simulations with Murrili falling teardrop-orientated (low drag), and then flat-orientated
352  (high drag). We indicate these directions of travel shown arrows $V_L$ and $V_H$ in Figure 9). In
353  Figure 10 we show the results of these Monte Carlo simulations. We exactly specify the
354  meteorite mass and density, but permit initial vector variation and 5% wind uncertainties. For
355  reference to previous figures we also display the 00:00 fall line (based on a spherical drag
356  coefficient, but with the later WRF v3.9 wind model) and meteorite recovery location.

357
358

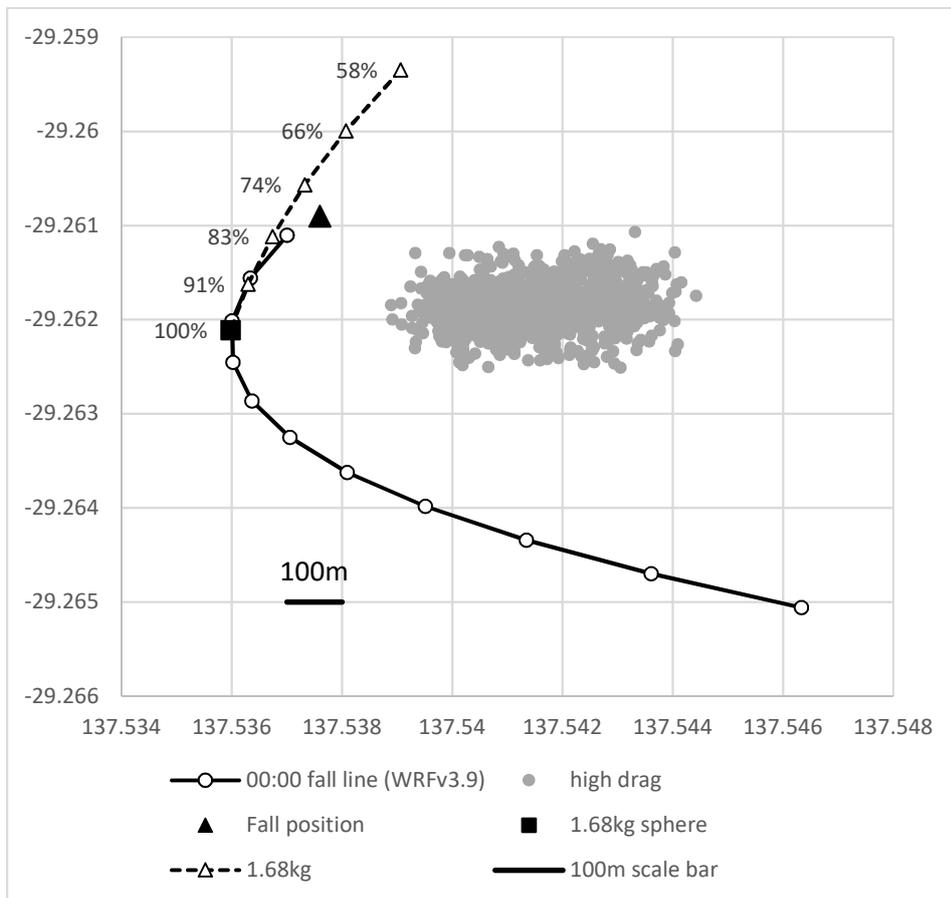

359
360

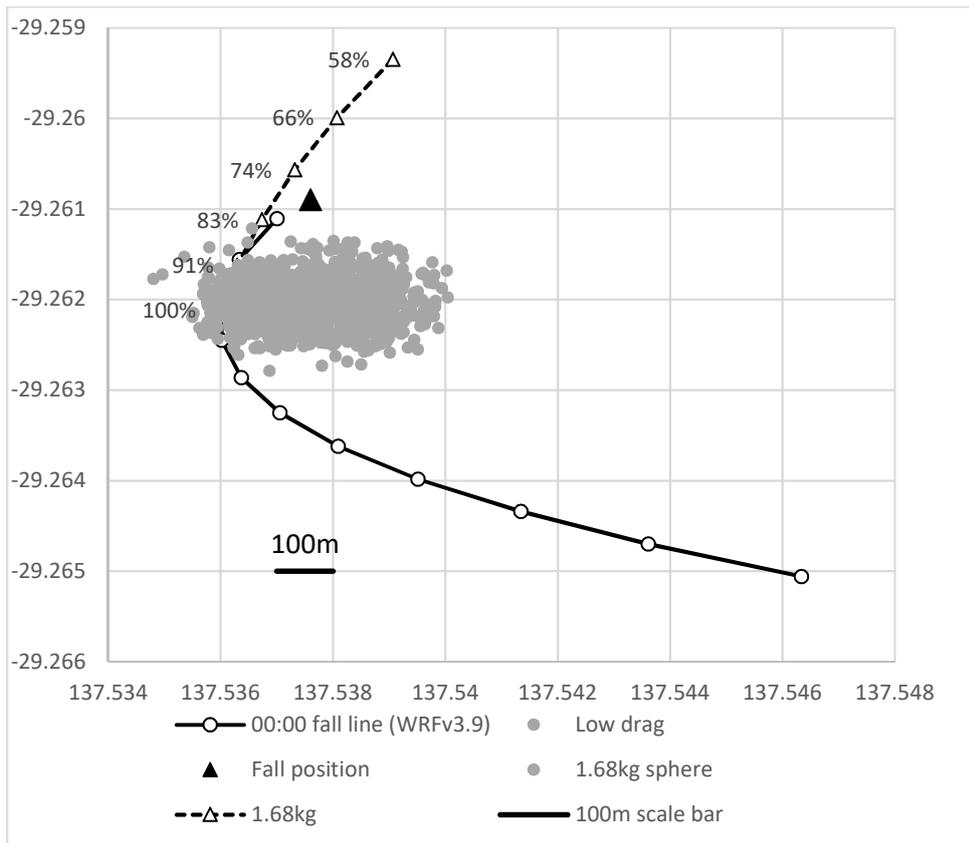

Considering the scatter points in Figure 10, we can see that fall orientation clearly will have an effect. (a) shows that the high drag orientated simulations are somewhat offset from the isotropic drag fall line, and from the recovery position (triangular marker). Figure 10(b) shows the vertically orientated (falling in a teardrop orientation, with lowest possible drag coefficient in direction of travel) scatter as a closer match, intersecting the isotropic fall line but not intersecting the recovery position

The actual position of the impact is close to the isotropic fall line, but ~400 m along the line from the 1.68 kg prediction (black square marker) which corresponds to an anomalously large mass (>5 kg). By beginning with a spherical isotropic object, and keeping mass fixed at 1.68 kg, but reducing the drag coefficient in all directions (so using Table 2, but with a scalar reduction in forces as labelled), we generate the dashed/triangle line that is oblique to the isotropic fall line. This line also doesn't pass precisely through the impact point, and the closest approach is when drag reduced by about 75% (so a scalar factor of 0.75) compared to a sphere. The closest approach distance is 44 m.

380 The relative lack of cross-track offset between the impact point and the isotropic sphere fall
381 line may result from the wind modelling not matching reality: In general isotropic drag changes
382 from shape, density, or mass variations act to effectively shift objects either along the fall line
383 or very close to it, rather than away from the line (The sphere and brick fall lines in Figure 2
384 are essentially the same line extended). Note also how the isotropic sub-sphere line in Figure
385 10 sub-parallel to the sphere line, not at some arbitrary angle. The recovered meteorite does
386 not precisely align with either of the isotropic lines; this offset is either from small amount of
387 non-spherical drag (aerodynamic shape effects such as angle of attack or lift), or most likely
388 from just basic inaccuracies in the wind model.

389 Comparing the Monte Carlo point clouds for the two orientations, the two orientated point
390 clouds do not bracket the spherical fall line, as one might initially imagine, as the influence of
391 sphericity means that sideways wind effects are decoupled from vertical velocity. The
392 difference between the centres of the clouds is shifted East and slightly South (from high to
393 low drag), which corresponds to the general wind orientations of approximately 270-300° in
394 Figure 5b. These point clouds can be thought of as representing extremes on a spectrum of
395 orientations, and some intermediate orientation value would lie in between these clouds.
396 However, no intermediate value would pass through the actual impact site, which hints that
397 perhaps the modelled wind data has some inaccuracy, most likely due to lack of real
398 supporting measurements.

399 In both cases the diameter and scatter of the orientated Monte Carlo point clouds are smaller
400 than the isotropic sphere scatter as seen in Figure 8b: We hypothesis that the sphere has a
401 unique combination of properties of wind influence and vertical fall velocity; so, the low drag
402 orientation is most influenced by the horizontal winds, but vertically falls a lot quicker, with less
403 time for the wind influence to act, and vice versa for the high drag orientation.

404 In Figure 10a, the difference in position between the high drag orientated point cloud and the
405 isotropic sphere fall line (and the impact point) appears relatively large, compared to the low
406 drag cloud. This appears somewhat counterintuitive, as one might expect the low drag down
407 orientation (which has a high drag sideways) to have bigger shift from the sphere line. This
408 shift must result from the higher drag giving a longer dwell time (slower fall velocity), giving
409 the winds greater influence.

410 Although the low drag (vertical orientation) Monte Carlo simulation appear closer than the high
411 drag simulation to the recovery point, neither orientation overlaps this point. In Figure 8b, the
412 Monte Carlo simulation of an isotropic sphere does however overlap the recovery point: The
413 wind model can be compatible with an isotropic object, but not an orientated fall, although an
414 irregular shape with complex tumbling may effectively cancel out any orientation effects.

415 So we result in several possible scenarios to reconcile the data with the modelling: wind
416 models approximately correct, but with an isotropic shape having an anomalously low drag

417 coefficient, or an isotropic shape where winds are lower or less influential than expected (or a
418 reduced drag coefficient). Finally, an orientated shape falling seems less likely, as the wind
419 models would have to be substantially incorrect, which is not supported by other recoveries,
420 both DFN and other fireball networks around the world.

421 These possibilities are not exclusive, as the Australian outback real wind observations are
422 sparse. A change of orientation during flight, and even rotation could not be ruled out, and
423 might have the effect of damping the effects of orientation. Even in the low drag scenario,
424 strong horizontal winds would have the effect of altering the angle of attack of the falling rock,
425 which will have aerodynamic effects. From this single event it is difficult to separate these
426 possibilities, but analysis of further falls should show which is the appropriate formulation to
427 use in future predictions.

428 These scatter plots, and also the shape dependence of points along the fall line in Figure 7
429 illustrate the both the need to search widely along an ideal fall line prediction, and the
430 dominance of shape in darkflight modelling. Any shape characteristics available from bright
431 flight behaviour will be most helpful, but detailed shape is unlikely to be known. The reasonable
432 match between basic predictions and the recovered fall position, would tend to indicate that
433 the wind model chosen in this case is probably sufficiently accurate (i.e. not the major source
434 of uncertainty), and that also that geometric errors in triangulation are relatively minor.
435 Assuming a specific orientation during darkflight does not provide best fit, but simple
436 assumptions can be helpful in planning searching. In contrast to the spherical or lower drag
437 coefficient here demonstrated, another DFN recovery, the Dingle Dell meteorite, landed 105
438 m from fall line, but at a point along the line that corresponds to a cylindrical mass, with a drag
439 coefficient significantly greater than spherical. ((Devillepoix et al. 2018), their figure 10).
440 However, we must note that Dingle Dell suffered from several complicating issues: the
441 recovered meteorite has an angular broken surface, implying fragmentation, which was
442 supported by the light curve; the entry angle was also shallow, and end point slightly higher
443 (19.1 km). However, in both Murrili and Dingle Dell cases the offset between fall lines and
444 recoveries, and the position along the line are relatively manageable a searching point of view.

445 ## 4 CONCLUSIONS

446 For fireball camera networks, focussed towards meteorite recovery, the calculation of the
447 darkflight trajectory after luminous observations is a critical step to sample recovery. This step
448 is very difficult to test, due to the lack of observations during descent, with only the recovery
449 (or not) of the meteorite providing a data point. The principles of darkflight calculation are
450 simple, based on a classical aerodynamic drag equation, but the calculation hides subtleties,
451 particularly in the formulation of aerodynamic properties. We describe the details and approach taken to this problem by the Desert Fireball Network. The simplest output from a darkflight calculation is typically a fall line, showing impact positions for a known range of hypothetical masses. From consideration of these lines, and the effects of parameters, we find that the choice of meteorite shape is more important than density or mass choice, in terms of variation in ground position. These constraints in turn influence the ground searching strategy. We illustrate this with a case study of the Murrili meteorite fall, recovered from Lake Eyre-Kati Thanda in 2015. Murrili is ideal case for darkflight study, as the optical observations and triangulation data were exceptionally good, with multiple DFN observatories relatively close by giving a range of viewpoints. Additionally, the meteor trajectory was steep, and the final height at the end of the luminous phase was at a relatively low altitude of 18 km, so the darkflight was relatively short, compared to many meteorite falls. However, winds were quite strong at this location, especially around the 15 km levels, so the darkflight fall line was perturbed significantly. Given the known location of the meteorite impact point and the known shape, we investigate whether the meteorite had an orientation during fall, andfind that although the final position can be matched using an orientation with the lowest drag coefficient in the direction of travel rather than the highest, the fall position is best matched by assuming a either a spherical shape and drag characteristics, or a reduced drag sphere, where one assumes spherical properties, and then artificially reduces the influence of atmospheric interactions (to about 75% in this case). A simple isotropic approach like this may provide a way forward to investigate weakly constrained shapes for observed falls; other falls seen by the DFN are also well matched with an isotropic shape, but not necessarily a pure sphere. We note from this that Murrili although is close to an ideal case for darkflight modelling, it was still necessary to consider in detail the overall shape of the meteorite and the detailed atmospheric properties in order to get a good agreement between predicted and observed fall positions.

Further work would benefit greatly from detailed published data concerning the shape of recovered meteorites, in combination with precise details of the end of luminous trajectory, so that aerodynamically realistic drag coefficients could be estimated, and compared to the recovery positions.

## 5 ACKNOWLEDGEMENTS

We thank the Arabana people, the traditional custodians of Kati Thanda, for their support during meteorite searching. This work is funded by the Australian Research as part of the Australian Discovery Project scheme (DP170102529). This work was supported by resources provided by the Pawsey Supercomputing Centre with funding from the Australian Government and the Government of Western Australia. We thank the Australian Wildlife Conservancy for

486 their support in maintaining the Kalamurina DFN camera. We thanks two anonymous
487 reviewers for the comments which greatly improved this paper.

# 6 REFERENCES

489 Astropy Collaboration, Price-Whelan AM, Sipőcz BM, et al (2018) The Astropy Project: Building an
490     Open-science Project and Status of the v2.0 Core Package. Astron J 156:123.
491     https://doi.org/10.3847/1538-3881/aabc4f

492 Borovicka J (1990) The comparison of two methods of determining meteor trajectories from
493     photographs. Bull Astron Inst Czechoslov 41:391–396

494 Borovička J, Kalenda P (2003) The Morávka meteorite fall: 4. Meteoroid dynamics and
495     fragmentation in the atmosphere. Meteorit Planet Sci 38:1023–1043.
496     https://doi.org/10.1111/j.1945-5100.2003.tb00296.x

497 Carter RTJ (2011) Constraining the Drag Coefficients of Meteors in Dark Flight

498 Ceplecha Z (1961) Multiple fall of Pribram meteorites photographed. Bull Ast Inst Cz 2:21

499 Ceplecha Z (1987) Geometric, dynamic, orbital and photometric data on meteoroids from
500     photographic fireball networks. Bull Astron Inst Czechoslov 38:222–234

501 Ceplecha Z, Borovicka J, Elford WG, et al (1998) Meteor phenomena and bodies. Space Sci Rev
502     84:327–471

503 Ceplecha Z, Borovička J, Spurný P (2000) Dynamical behavior of meteoroids in the atmosphere
504     derived from very precise photographic records. Astron Astrophys 357:1115–1122

505 Colas F, Zanda B, Bouley S, et al (2014) The FRIPON and Vigie-Ciel networks. pp 34–38

506 Connolly BJ, Loth E, Smith CF (2020) Shape and drag of irregular angular particles and test dust.
507     Powder Technol 363:275–285. https://doi.org/10.1016/j.powtec.2019.12.045

508 Corey AT (1949) Influence of shape on the fall velocity of sand grains. Colorado State University

509 Devillepoix HAR, Sansom EK, Bland PA, et al (2018) The Dingle Dell meteorite: A Halloween treat
510     from the Main Belt. Meteorit Planet Sci 53:2212–2227. https://doi.org/10.1111/maps.13142

511 Flynn GJ (2005) Physical Properties of Meteorites and Interplanetary Dust Particles: Clues to the
512     Properties of the Meteors and their Parent Bodies. In: Hawkes R, Mann I, Brown P (eds)
513     Modern Meteor Science An Interdisciplinary View. Springer Netherlands, pp 361–374

514 Folinsbee RE, Bayrock LA (1961) The Bruderheim Meteorite-Fall and Recovery. J R Astron Soc
515     Can 55:218

516 Fries M, Fries J (2010) Doppler weather radar as a meteorite recovery tool. Meteorit Planet Sci
517     45:1476–1487. https://doi.org/10.1111/j.1945-5100.2010.01115.x

518 Gritsevich M, Lyytinen E, Moilanen J, et al (2014) First meteorite recovery based on observations
519     by the Finnish Fireball Network

520 Gritsevich MI (2007) Approximation of the observed motion of bolides by the analytical solution of
521     the equations of meteor physics. Sol Syst Res 41:509–514.
522     https://doi.org/10.1134/S003809460706007X

523 Haider A, Levenspiel O (1989) Drag coefficient and terminal velocity of spherical and nonspherical
524     particles. Powder Technol 58:63–70. https://doi.org/10.1016/0032-5910(89)80008-7

525 Halliday I, Blackwell AT, Griffin AA (1978) The Innisfree meteorite and the Canadian Camera
526     Network. J R Ast Soc Can 72:15–39

527 Hölzer A, Sommerfeld M (2008) New simple correlation formula for the drag coefficient of non-
528     spherical particles. Powder Technol 184:361–365.
529     https://doi.org/10.1016/j.powtec.2007.08.021

530 Howie RM, Paxman J, Bland PA, et al (2017) How to build a continental scale fireball camera
531     network. Exp Astron 43:237–266. https://doi.org/10.1007/s10686-017-9532-7

532 Jacchia LG, Whipple FL (1956) The Harvard photographic meteor programme. Vistas Astron
533     2:982–994. https://doi.org/10.1016/0083-6656(56)90021-6

534 Jenniskens P, Fries MD, Yin Q-Z, et al (2012) Radar-Enabled Recovery of the Sutter's Mill
535     Meteorite, a Carbonaceous Chondrite Regolith Breccia. Science 338:1583–1587.
536     https://doi.org/10.1126/science.1227163

537 Jones E, Oliphant T, Peterson P (2001) SciPy: Open source scientific tools for Python

538 Khanukaeva DYu (2003) On the Coefficients in Meteor Physics Equations. AIP Conf Proc
539     663:726–734. https://doi.org/10.1063/1.1581615

Kleinstreuer C, Feng Y (2013) Computational Analysis of Non-Spherical Particle Transport and Deposition in Shear Flow With Application to Lung Aerosol Dynamics—A Review. J Biomech Eng 135:. https://doi.org/10.1115/1.4023236

Masson DJ, Morris DN, Bloxom DE (1960) Measurements of sphere drag from hypersonic continuum to free-molecule flow. RAND Corporation

McCrosky RE, Posen A, Schwartz G, Shao C-Y (1971) Lost City meteorite—Its recovery and a comparison with other fireballs. J Geophys Res 76:4090–4108. https://doi.org/10.1029/JB076i017p04090

Miller DG, Bailey AB (1979) Sphere drag at Mach numbers from 0·3 to 2·0 at Reynolds numbers approaching 10^7. J Fluid Mech 93:449–464. https://doi.org/10.1017/S0022112079002597

Passey QR, Melosh HJ (1980) Effects of atmospheric breakup on crater field formation. Icarus 42:211–233. https://doi.org/10.1016/0019-1035(80)90072-X

Pecina P, Ceplecha Z (1983) New aspects in single-body meteor physics. Bull Astron Inst Czechoslov 34:102–121

Pettijohn FJ (1975) Sedimentary Rocks. Harper and Row, New York

ReVelle DO (2005) Recent Advances in Bolide Entry Modeling:A Bolide Potpourri*. Earth Moon Planets 97:1–35. https://doi.org/10.1007/s11038-005-2876-4

Revelle DO (2002) Fireball dynamics, energetics, ablation, luminosity and fragmentation modeling. pp 127–136

Sansom EK, Bland P, Paxman J, Towner M (2015) A novel approach to fireball modeling: The observable and the calculated. Meteorit Planet Sci 50:1423–1435. https://doi.org/10.1111/maps.12478

Sansom EK, Bland PA, Towner MC, et al (2020) Murrili meteorite's fall and recovery from Kati Thanda. Meteorit Planet Sci 55:2157–2168. https://doi.org/10.1111/maps.13566

Sansom EK, Gritsevich M, Devillepoix HAR, et al (2019) Determining Fireball Fates Using the alpha–beta Criterion. Astrophys J 885:115. https://doi.org/10.3847/1538-4357/ab4516

Sansom EK, Rutten MG, Bland PA (2017) Analyzing Meteoroid Flights Using Particle Filters. Astron J 153:87. https://doi.org/10.3847/1538-3881/153/2/87

Skamarock C, Klemp B, Dudhia J, et al (2019) A Description of the Advanced Research WRF Model Version 4. https://doi.org/10.5065/1dfh-6p97

Skamarock WC, Klemp JB, Dudhia J, et al (2008) A Description of the Advanced Research WRF Version 3. National Center for Atmospheric Researc

Spurny P, Bland PA, Borovicka J, et al (2012) The Mason Gully Meteorite Fall in SW Australia: Fireball Trajectory, Luminosity, Dynamics, Orbit and Impact Position from Photographic Records. 1667:6369

Spurný P, Bland PA, Shrbený L, et al (2012) The Bunburra Rockhole meteorite fall in SW Australia: fireball trajectory, luminosity, dynamics, orbit, and impact position from photographic and photoelectric records. Meteorit Planet Sci 47:163–185. https://doi.org/10.1111/j.1945-5100.2011.01321.x

Spurný P, Borovička J, Shrbený L (2006) Automation of the Czech part of the European fireball network: equipment, methods and first results. Proc Int Astron Union 2:121–130. https://doi.org/10.1017/S1743921307003146

Turchak LI, Gritsevich MI (2014) Meteoroids Interaction With The Earth Atmosphere. J Theor Appl Mech 44:15–28. https://doi.org/10.2478/jtam-2014-0020

Vinnikov VV, Gritsevich MI, Turchak LI (2016) Mathematical model for estimation of meteoroid dark flight trajectory. AIP Conf Proc 1773:110016. https://doi.org/10.1063/1.4965020

Wadell H (1935) Volume, Shape, and Roundness of Quartz Particles. J Geol 43:250–280. https://doi.org/10.1086/624298

Young CW (1967) Development Of Empirical Equations For Predicting Depth Of An Earth-Penetrating Projectile. Sandia Corp., Albuquerque, N. Mex.

Young CW (1997) Penetration equations. Sandia National Lab. (SNL-NM), Albuquerque, NM (United States)

Zhdan IA, Stulov VP, Stulov PV, Turchak LI (2007) Drag coefficients for bodies of meteorite-like shapes. Sol Syst Res 41:505–508. https://doi.org/10.1134/S0038094607060068